\documentclass[aps,prl,reprint,superscriptaddress]{revtex4-1}

\usepackage{graphicx}
\usepackage{amsmath}
\usepackage{amsfonts}
\usepackage{amssymb}
\usepackage{color}

\newcounter{defcounter}
\newcommand{\ub}{\overline{u}}

\newcommand{\sgn}{\mathrm{sgn}}

\begin{document}

\title{Solitonic dispersive hydrodynamics: theory and observation} 

\author{Michelle D. Maiden}
\author{Dalton V. Anderson}
\author{Nevil A. Franco}
\affiliation{Department of Applied Mathematics, University of Colorado, Boulder, Colorado 80309-0526, USA}
\author{Gennady A. El}
\affiliation{Department of Mathematical Sciences, Loughborough University,
  Loughborough, LE11 3TU, UK}
\author{Mark A. Hoefer} 
\email{hoefer@colorado.edu}
\affiliation{Department of Applied Mathematics, University of
  Colorado, Boulder, Colorado 80309-0526, USA}
\date{\today}


\begin{abstract}
  Ubiquitous nonlinear waves in dispersive media include localized
  solitons and extended hydrodynamic states such as dispersive shock
  waves.  Despite their physical prominence and the development of
  thorough theoretical and experimental investigations of each
  separately, experiments and a unified theory of solitons and
  dispersive hydrodynamics are lacking.  Here, a general soliton-mean
  field theory is introduced and used to describe the propagation of
  solitons in macroscopic hydrodynamic flows.  Two universal adiabatic
  invariants of motion are identified that predict trapping or
  transmission of solitons by hydrodynamic states.  The result of
  solitons incident upon smooth expansion waves or compressive,
  rapidly oscillating dispersive shock waves is the same, an effect
  termed hydrodynamic reciprocity.  Experiments on viscous fluid
  conduits quantitatively confirm the soliton-mean field theory with
  broader implications for nonlinear optics, superfluids, geophysical
  fluids, and other dispersive hydrodynamic media.
\end{abstract}

\maketitle

Long wavelength, hydrodynamic theories abound in physics, from fluids
\cite{landau_fluid_1987} to optics \cite{carusotto_quantum_2013},
condensed matter \cite{fradkin_field_2013} to quantum mechanics
\cite{wyatt_quantum_2005}, and beyond.  Such theories describe
expansion and compression waves until breaking.  When the physics at
shorter wavelengths are predominantly dispersive, dispersive
hydrodynamic theories \cite{whitham_linear_1974,el_dispersive_2016}
are used to describe shock waves of a spectacularly different
character than their dissipative counterparts.  Dispersive shock waves
(DSWs) consist of coherent, rank-ordered, nonlinear oscillations that
continually expand
\cite{gurevich_nonstationary_1974,el_dispersive_2016}.  Observations
in a wide range of physical media that include quantum matter
\cite{hoefer_dispersive_2006-1,mo_experimental_2013}, optics
\cite{wan_dispersive_2007,xu_dispersive_2017-1}, classical fluids
\cite{trillo_observation_2016,maiden_observation_2016} and magnetic
materials \cite{janantha_observation_2017} demonstrate the prevalence
of DSWs.

Another celebrated feature of dispersive hydrodynamic media are
localized, nonlinear solitary waves.
When they exhibit particle-like properties such as elastic, pairwise
interactions, solitary waves are called solitons
\cite{zabusky_interaction_1965} and have been extensively studied both
theoretically \cite{drazin_solitons:_1989} and experimentally
\cite{remoissenet_waves_2013}.  The focus here is on solitary waves
that exhibit solitonic behavior, i.e., elastic or near-elastic
interaction, henceforth we refer to them as solitons.  Despite their
common origins, solitons and dispersive hydrodynamics have been
primarily studied independently.

\begin{figure}
  \centering
  \includegraphics{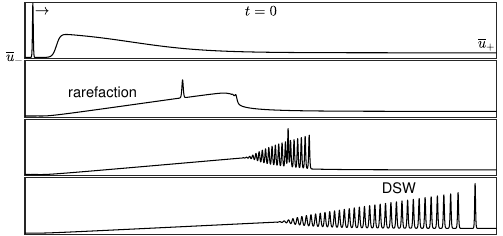}
  \caption{Representative initial configuration and evolution (top to
    bottom) for solitonic dispersive hydrodynamics.  The narrow
    soliton on the uniform mean field $\overline{u}_-$ is transmitted
    through the broad hydrodynamic flow if it reaches and propagates
    freely on the uniform mean field $\overline{u}_+$.  The
    hydrodynamic flow exhibits expansion (rarefaction) and compression
    that leads to a dispersive shock wave.}
  \label{fig:config}
\end{figure}

Utilizing the scale separation between extended hydrodynamic states
and localized solitons (see Fig.~\ref{fig:config}), we propose in this
work a general theory of solitonic dispersive hydrodynamics
encapsulated by a set of effective partial differential equations for
the hydrodynamic mean field, the soliton's amplitude, and its phase.
We identify two adiabatic invariants of motion and show that they lead
to two pivotal predictions.  First, the soliton trajectory is a
characteristic of the governing equations that is directed by the mean
field, a nonlinear analogue of wavepacket trajectories in quantum
mechanics \cite{wyatt_quantum_2005}.  This implies that solitons are
either trapped by or transmitted through a hydrodynamic state,
depending on the relative amplitudes of the soliton and the
hydrodynamic ``barrier''.

The second prediction we term hydrodynamic reciprocity.  Given an
incident soliton amplitude and the far-field mean conditions, the
adiabatic invariants are used to predict when the soliton is trapped
or transmitted and, in the latter case, what its transmitted amplitude
and phase shift are.  Hydrodynamic reciprocity means that the
trapping, transmission amplitude/phase relations are the same for
soliton interactions with smooth, expanding rarefaction waves (RWs)
and compressive, oscillatory DSWs.

\begin{figure*}
  \centering
  \includegraphics[width=18cm]{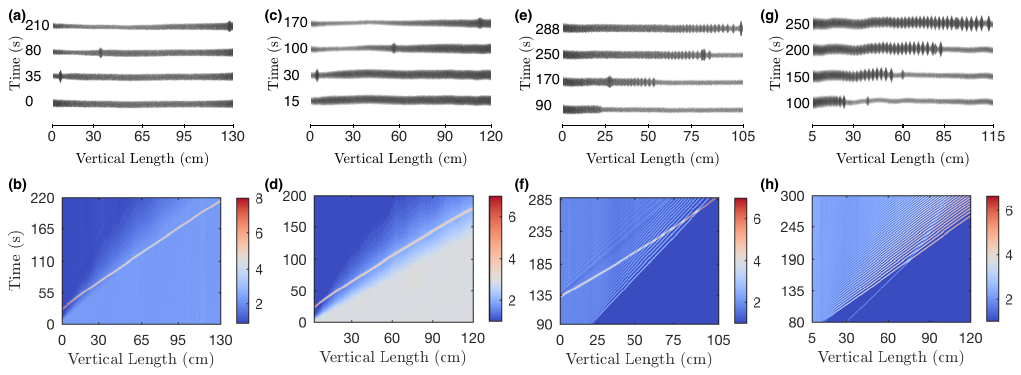}
  \caption{Experiments demonstrating soliton transmission and trapping
    with hydrodynamic states.  Representative image sequences
    (a,c,e,g) and space-time contours (b,d,f,h) extracted from image
    processing are shown.  The contour intensity scale is the
    dimensionless conduit cross-sectional area relative to the
    smallest area. a,b) Soliton-RW transmission.  c,d) Soliton-RW
    trapping.  e,f) Soliton-DSW transmission.  g,h) Soliton-DSW
    trapping.}
  \label{fig:expt}
\end{figure*}
We confirm these predictions with experiments on the interfacial
dynamics of a viscous fluid conduit, a model dispersive hydrodynamic
medium \cite{lowman_dispersive_2013-1} that has been used previously
to investigate solitons
\cite{olson_solitary_1986,scott_observations_1986,lowman_interactions_2014}
and DSWs \cite{maiden_observation_2016}. Although soliton-DSW
interaction has been observed previously
\cite{maiden_observation_2016}, the nature and properties of the
interaction were not explained.  We stress that the theory presented
is general and applies to a wide range of physical media
\cite{hoefer_dispersive_2006-1,wan_dispersive_2007,mo_experimental_2013,maiden_observation_2016,trillo_observation_2016,xu_dispersive_2017-1,janantha_observation_2017}.

Experiments are performed on the interfacial dynamics of a buoyant,
viscous fluid injected from below into a miscible, much more viscous
fluid matrix.  Due to negligible diffusion and high viscosity
contrast, the two-fluid interface serves as the dispersive
hydrodynamic medium
\cite{olson_solitary_1986,lowman_dispersive_2013-1}.  The experimental
setup is similar to that described in \cite{maiden_observation_2016}
and consists of a tall acrylic column
filled with glycerol (viscosity $1.2 \pm 0.2$ P, density $1.2587 \pm
0.0001$ g/cm$^3$).  A nozzle at the column base serves as the
injection point for the interior fluid (viscosity $0.51\pm 0.01$ P,
density $1.2286 \pm 0.0001$ g/cm$^3$), a miscible solution of
glycerol, water, and black food coloring.  By injecting at a constant
rate (0.25 mL/min or 0.77 mL/min), the buoyant interior fluid
establishes a vertically uniform fluid conduit.  Although predicted to
be unstable, our experiment operates in the convective regime
\cite{selvam_convective_2009}.  By varying the injection rate, conduit
solitons, RWs, and DSWs can be generated at the interface between the
interior and exterior fluids.

Observations of the hydrodynamic transmission and trapping of solitons
resulting from their interaction with RWs and DSWs are depicted in
Fig.~\ref{fig:expt}.  The contour plots in \ref{fig:expt}(b,f) show
that transmitted solitons exhibit a smaller (larger) amplitude and
faster (slower) speed post interaction with a RW (DSW).  The
transmitted solitons experience a phase shift due to hydrodynamic
interaction, defined as the difference between the post and pre
interaction spatial intercept.  Our measurements show a negative
(positive) phase shift for the soliton transmitted through a RW (DSW).
Sufficiently small incident solitons in Fig.~\ref{fig:expt}(d,h) do
not emerge from the RW or DSW interior during the course of
experiment, remaining trapped inside the hydrodynamic state.

We now present a theory to explain these observations by considering a
general dispersive hydrodynamic medium with nondimensional scalar
quantity $u(x,t)$ (e.g., conduit cross-sectional area) governed by
\begin{equation}
  \label{eq:1}
  u_t + V(u) u_x = D[u]_x, \quad x \in \mathbb{R}, \quad t > 0 .
\end{equation}
$V(u)$ is the long-wave speed, $D[u]$ is an integro-differential
operator, and Eq.~(\ref{eq:1}) admits a real-valued, linear dispersion
relation with frequency $\omega(k,\ub)$
where $k$ is the wavenumber and $\overline{u}$ is the background mean
field.  We assume $V'(u) > 0$ so that the dispersive hydrodynamic
system has convex flux \cite{el_dispersive_2017}. The dispersion
is assumed negative ($\omega_{kk} < 0$) for definiteness.  We also
assume that equation \eqref{eq:1} satisfies the prerequisites for
Whitham theory, an approximate description of modulated nonlinear
waves that accurately characterizes dispersive hydrodynamics in a
wide-range of physical systems
\cite{whitham_linear_1974,el_dispersive_2016}.

Many models can be expressed in the form \eqref{eq:1}.  In the
Appendix, we perform calculations for
the Korteweg-de Vries (KdV) equation $V(u) = u$, $D[u] = -u_{xx}$, a
universal model of weakly nonlinear, dispersive waves, and the conduit
equation $V(u) = 2u$, $D[u] = u^2(u^{-1}u_t)_x$, an accurate model for
our experiments \cite{lowman_dispersive_2013-1}.

The dynamics of DSWs, RWs, and solitons for Eq.~(\ref{eq:1}) can be
described using Whitham theory \cite{whitham_linear_1974}, where a
nonlinear periodic wave's mean $\overline{u}$, amplitude $a$, and
wavenumber $k$ are assumed to vary slowly via modulation equations.
The modulation equations admit an asymptotic reduction in the
non-interacting soliton wavetrain regime $0 < k \ll 1$
\cite{grimshaw_slowly_1979,gurevich_nonlinear_1990}
\begin{equation}
  \label{eq:3}
  \begin{split}
    &\overline{u}_t + V(\overline{u})\overline{u}_x = 0 , \quad a_t +
    c(a,\overline{u}) a_x + f(a,\overline{u}) \overline{u}_x = 0, \\
    &k_t + \left [ c(a,\ub \right) k ]_x = 0 .
  \end{split}
\end{equation}
The first equation is for the decoupled mean field, which is governed
by the dispersionless, $D \to 0$, equation \eqref{eq:1}. The second
equation describes the soliton amplitude $a$, which is advected by the
mean field according to the soliton amplitude-speed relation
$c(a,\overline{u})$ and the coupling function $f(a,\overline{u})$.
The final equation expresses wave conservation
\cite{whitham_linear_1974} and describes a train of solitons with
spacing $2\pi/k \gg 1$.  The soliton train here is a useful, yet
fictitious construct because we will only consider the soliton limit
$k \to 0$ of solutions to Eq.~(2).  Equation~\eqref{eq:3} with
$c(a,\overline{u}) = a/3 + \overline{u}$ and $f(a,\overline{u}) =
2a/3$ corresponds to the soliton limit of the KdV-Whitham system of
modulation equations, shown in \cite{el_evolution_2007} to be
equivalent to the soliton modulation equations determined by other
means \cite{grimshaw_slowly_1979} with application to shallow water
soliton propagation over topography in
\cite{grimshaw_slowly_1979,el_transformation_2012,grimshaw_propagation_2016,grimshaw_depression_2016}.
The general case of Eq.~(\ref{eq:3}) was derived in
\cite{gurevich_nonlinear_1990} and can be interpreted as a mean field
approximation for the interaction of a soliton with the hydrodynamic
flow.  In contrast to standard soliton perturbation theory where the
soliton's parameters evolve temporally \cite{kivshar_dynamics_1989},
solitonic dispersive hydrodynamics require the soliton amplitude
$a(x,t)$ be treated as a spatio-temporal field.  We note that the
equations in \eqref{eq:3} can be solved sequentially by the method of
characteristics \cite{grimshaw_slowly_1979}.

It will be physically revealing to diagonalize the system of equations
in \eqref{eq:3} by identifying its Riemann invariants
\cite{whitham_linear_1974}.  Owing to the special structure of
(\ref{eq:3}) with just two characteristic velocities $V<c$, it is
always possible to find a change of variables to Riemann invariant
form that diagonalizes the system.  The mean field equation is already
diagonalized with $\ub$ the Riemann invariant associated to the
velocity $V$.  The second Riemann invariant, $q = q(a,\ub)$ is
associated with the velocity $c$.  $q$ can be found by integrating the
differential form $f \mathrm{d}\ub + (c-V) \mathrm{d} a$
\cite{whitham_linear_1974} (see the Appendix).  For KdV, $q(a,\ub) =
a/2 + \ub$, whereas for the conduit equation
\begin{equation}
  \label{eq:8}
  \begin{split}
   c(a,\ub) &= [\ub^2 + (a+\ub)^2(2\ln(1+a/\ub)-1)]\ub/a^2, \\
   q(a,\ub) &= c(a,\ub)[c(a,\ub)+2\ub]/\ub .
  \end{split}
\end{equation}
The third Riemann invariant is found by direct integration of the
wavenumber equation to be the quantity $k p(q,\ub)$ given by
\begin{equation}
  \label{eq:4}
     p(q,\ub)  = \exp \left (- \int_{\ub_0}^{\ub}
    \frac{C_{u}(q,u)}{V(u) - C(q,u)} \,\mathrm{d}u \right),
\end{equation}
where $C(q(a,\ub), \ub) \equiv c(a, \ub)$.  For KdV,
$p(q,\ub) = (q-\ub)^{-1/2}$.  The change of variables $q = q(a,\ub)$
and $p = p(q,\ub)$ diagonalizes (\ref{eq:3})
\begin{equation}
  \label{eq:30}
  \begin{split}
    &\ub_t + V(\ub)\ub_x = 0, \quad q_t +  C(q,\ub) q_x = 0 ,\\
    &(kp)_t +  C(q,\ub) (kp)_x = 0 .
  \end{split}
\end{equation}

We seek solutions to Eq.~\eqref{eq:30} subject to an initial mean
field profile $\ub(x,0) = \ub_0(x)$ and an initial soliton of
amplitude $a_0$ located at $x = x_0$.  But we require initial soliton
and wavenumber fields $a(x,0)$ and $k(x,0)$ for all $x$ in order to
give a properly posed problem for \eqref{eq:3}.  Admissible initial
conditions are obtained by recognizing this as a special solution, a
simple wave in which all but one of the Riemann invariants are
constant \cite{whitham_linear_1974}.  The non-constant Riemann
invariant must be $\ub$ to satisfy the initial condition and therefore
satisfies $\ub = \ub_0(x - V(\ub)t)$.  The initial soliton amplitude
and position determine the constant Riemann invariant $q_0 =
q(a_0,\ub_0(x_0))$.  An initial wavenumber $k_0$ determines the other
constant Riemann invariant $k_0p_0 = k_0p(q_0,\ub_0(x_0))$.  As we
will show, the value of $k_0 > 0$ is not relevant so can be
arbitrarily chosen.  We now show how this solution physically
describes soliton-mean field interaction.

A smooth, initial mean field, e.g., in Fig.~\ref{fig:config}, will
evolve according to the obtained implicit solution until wavebreaking
occurs.  Our interest is in the interaction of a soliton with the
expansion and compression waves that result.  In dispersive
hydrodynamics, the simplest examples of these are RWs and DSWs,
respectively, which are most conveniently generated from step initial
data.  We therefore analyze the obtained general solution subject to
step initial data
\begin{equation}
  \label{eq:7}
  \overline{u}(x,0) = \overline{u}_{\pm}, ~ a(x,0) = a_\pm, ~
  k(x,0) = k_\pm, ~ \pm x > 0 ,
\end{equation}
that model incident and transmitted soliton amplitudes $a_-$ and $a_+$
through the mean field transition $\ub_-$ to $\ub_+$ for soliton train
wavenumbers $k_-$ and $k_+$.  The mean field dynamics depend upon the
ordering of $\ub_-$ and $\ub_+$.  When $\ub_- < \ub_+$, the mean field
equation admits a RW solution, otherwise an unphysical, multi-valued
solution.  Short-wave dispersion regularizes such behavior and leads
to the generation of a DSW.  We consider each case in turn.

The transmission of a soliton through a RW is shown experimentally in
Fig.~\ref{fig:expt}(a,b).  The incident soliton ``climbs'' the RW and
emerges from the interaction with altered amplitude and speed.
The mean field is the self-similar, RW solution with
$u(x,t) = \ub_\pm$ for $\pm x > \pm V_\pm t$ and
\begin{equation}
  \label{eq:9}
  \overline{u}(x,t) = V^{-1}(x/t), \quad  V_-t \le x \le  V_+t,
\end{equation}
where $V_\pm = V(\ub_\pm)$ and $V^{-1}$ is the inverse of $V$.
Constant $q$ and $kp$ correspond to adiabatic invariants of the
soliton-mean field dynamics that yield constraints on the amplitude,
mean field, and wavenumber parameters we call the \emph{transmission
  and phase conditions}
\begin{equation}
  \label{eq:34}
  q(a_-,\ub_-) = q(a_+,\ub_+) , \quad \frac{k_-}{k_+} =
  \frac{p(q_+,\ub_+)}{p(q_-,\ub_-)} .
\end{equation}
The first adiabatic invariant $q(a,\ub)$ determines the transmitted
soliton amplitude $a_+$ in terms of the incident soliton amplitude
$a_-$ and the mean fields $\ub_\pm$.  The second adiabatic invariant
determines the ratio $k_-/k_+$, which in turn yields the soliton's
phase shift due to hydrodynamic interaction.  Equation (\ref{eq:34})
is the main theoretical result of this work and describes the
trapping or transmission of a soliton through a RW and a DSW.

The necessary and sufficient condition for soliton transmission is a
positive transmitted soliton amplitude $a_+ > 0$, which places a
restriction on the incident soliton amplitude $a_-$.  For the conduit
equation, Eq.~\eqref{eq:8} implies
$c_- > c_{\rm cr} = -\ub_- + (\ub_-^2 + 8\ub_+\ub_-)^{1/2}$.  For KdV,
$a_- > a_{\rm cr} = 2(\ub_+ - \ub_-)$. In both cases, we find that the
transmitted soliton's amplitude is decreased, $a_+ < a_-$ and its
speed is increased, $c_+ > c_-$.  More generally,
$\mathrm{sgn}(a_+-a_-) = -\mathrm{sgn}(q_{\ub}q_a)$ and
$\mathrm{sgn}(c_+-c_-) = \mathrm{sgn}(C_{\ub})$ (see Appendix).

The soliton phase shift is $\Delta = x_+ - x_-$ where $x_\pm$ are the
$x$-intercepts of the soliton pre ($-$) and post ($+$) hydrodynamic
interaction.  Given the initial soliton position $x_-$, the
contraction/expansion of the soliton train determines the phase shift
as $\Delta/x_- = k_-/k_+-1 = p_+/p_--1$.  Hence, the ratio $k_-/k_+$
in the phase condition \eqref{eq:34}, not the arbitrary initial
wavenumber $k_-$, determines the soliton phase shift.  Our use of a
fictitious soliton train is therefore justified.

We also determine the soliton-RW trajectory.  A soliton with position
$x(t)$ propagates through the mean field along a characteristic of
the modulation system (\ref{eq:3})
\begin{equation}
  \label{eq:10}
  \frac{dx}{dt} = C(q,\overline{u}(x,t)), \quad x(0) = x_- ,
\end{equation}
where the soliton amplitude $a(x,t)$ varies along the trajectory
according to the adiabatic invariant
$q(a(x,t),\overline{u}(x,t)) = q(a_-,\overline{u}_-)$.
The phase shift from integration of \eqref{eq:10} equals $\Delta$ from
the adiabatic invariant in \eqref{eq:34}, as expected.

When $a_+ \le 0$ in (\ref{eq:34}), the soliton is trapped by the RW,
as in experiment,
Fig.~\ref{fig:expt}(c,d).



If $\ub_- > \ub_+$, a DSW is generated. Soliton-DSW transmission is
experimentally depicted in Fig.~\ref{fig:expt}(e,f).  An incident
soliton propagates through the DSW, exhibiting a highly non-trivial
interaction, ultimately emerging with altered amplitude and speed.

\begin{figure}
  \centering
  \includegraphics{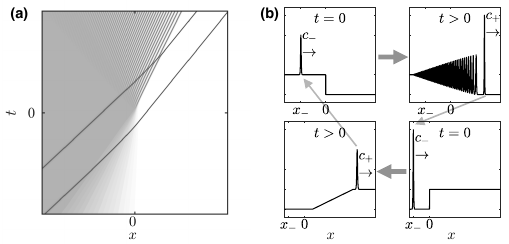}
  \caption{Graphical depictions of hydrodynamic reciprocity.  (a)
    Space-time contour plot of soliton-DSW ($t > 0$) and soliton-RW
    ($t < 0$) interaction with two solitons satisfying the
    transmission condition \eqref{eq:34}.  For $|t|$ sufficiently
    large, the soliton speeds are the same.  (b) If the soliton post
    DSW interaction (top, left to right) is used to initialize
    soliton-RW interaction (bottom, right to left), the post RW
    interaction soliton has the same properties as the pre DSW
    interaction soliton. }
  \label{fig:hydrodynamic_reciprocity}
\end{figure}
In contrast to the soliton-RW problem, the modulation equations
(\ref{eq:3}) are no longer valid throughout the soliton-DSW
interaction.  Instead, the mean field equation is replaced by the DSW
modulation equations
\cite{gurevich_nonstationary_1974,el_dispersive_2016}.  We seek a
simple wave solution for soliton-DSW modulation.  Because DSW
generation occurs only for $t > 0$, the soliton-DSW modulation system
for $t < 0$ reduces exactly to Eq.~(\ref{eq:3}), i.e., that of
soliton-RW modulation.  For $t < 0$, the adiabatic invariants
\eqref{eq:34} hold.  By continuity of the modulation solution, these
conditions must hold for $t \ge 0$ as well.  In particular, soliton-RW
and soliton-DSW interaction both satisfy \textit{the same transmission
  and phase conditions} (\ref{eq:34}).  This fact, termed hydrodynamic
reciprocity, is due to time reversibility of the governing equation
(\ref{eq:1}) and is depicted graphically in
Fig.~\ref{fig:hydrodynamic_reciprocity}.


Equations \eqref{eq:8} and \eqref{eq:34} for the conduit equation
indicate that solitons incident upon DSWs exhibit a decreased
transmitted speed $c_{\rm cr} < c_+ < c_-$ and an increased
transmitted amplitude $a_+ > a_{\rm cr} > a_-$.
$a_{\rm cr}$ and $c_{\rm cr}$ are precisely the amplitude and speed of
the DSW's soliton leading edge \cite{lowman_dispersive_2013}.
Hydrodynamic reciprocity therefore implies that the transmitted
soliton's amplitude is decreased (increased), its speed is increased
(decreased), and its phase shift is negative (positive) relative to
the soliton incident upon the RW (DSW), as observed experimentally in
Fig.~\ref{fig:expt}.  Using the transmission and phase conditions
\eqref{eq:34}, we accurately predict the conduit soliton trajectory
post DSW interaction without any detailed knowledge of soliton-DSW
interaction (see Appendix).

In contrast to soliton-RW transmission, solitons with amplitude $a_+$
initially placed to the right of the step will interact with the DSW
if $a_+ < a_{\rm cr}$.  Then the transmission condition \eqref{eq:34}
implies $a_- < 0$, i.e., the soliton cannot transmit back through the
DSW.  Instead, the soliton is effectively trapped as a localized
defect in the DSW interior as observed experimentally in
Fig.~\ref{fig:expt}(g,h).

\begin{figure}
  \centering
  \includegraphics{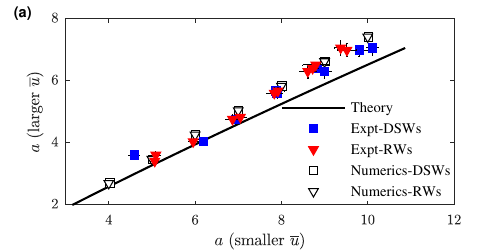}\\
  \includegraphics[scale=0.5]{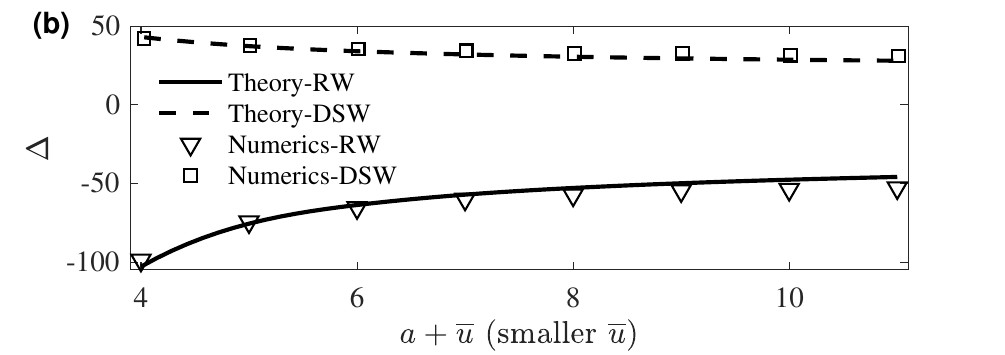}
  \caption{Transmitted soliton properties due to conduit soliton-RW
    and DSW interaction for a hydrodynamic transition from $\ub = 1$
    to $\ub = 1.75$. a) Soliton amplitude from eq.~\eqref{eq:34}
    (curve), experiment (filled squares, triangles), and numerical
    simulations (open squares, triangles). 
    b) Soliton phase shift from eq.~\eqref{eq:34} (curves) and
    numerical simulations (symbols).}
  \label{fig:transmission_condition}
\end{figure}
The transmission and phase conditions \eqref{eq:34} for the conduit
equation are shown in Fig.~\ref{fig:transmission_condition}.  For
soliton-RW interaction, the abscissa and ordinate are $a_-$ and $a_+$,
respectively reversed for soliton-DSW interaction.  Hydrodynamic
reciprocity implies that the transmission condition on these axes is
the same for soliton-RW and DSW transmission.  Reciprocity is
confirmed by experiment and numerical simulations of the conduit
equation in Fig.~\ref{fig:transmission_condition}(a), that slightly
deviate from soliton-mean field theory as the amplitudes increase,
consistent with previously observed discrepancies
\cite{lowman_dispersive_2013,maiden_observation_2016}.  Reciprocity of
the phase shift is also confirmed by conduit equation numerics in
Fig.~\ref{fig:transmission_condition}(b).  Our experiments provide
definitive evidence of soliton-hydrodynamic transmission, trapping,
reciprocity, and the theory's efficacy.

We have introduced a general framework for soliton-mean field
interaction.  The dynamics exhibit two adiabatic invariants that
describe soliton trapping or transmission.  The existence of the same
adiabatic invariants for soliton-mean field interactions of
compression (DSW) and expansion (RW) imply hydrodynamic reciprocity.
This describes a conceptually new notion of hydrodynamic soliton
``tunneling'' where the potential barrier is the mean field, obeying
the same equations as the soliton \cite{sprenger_soliton_2018}.



\appendix


\renewcommand{\theequation}{A\arabic{equation}}

\section{Appendix A:  Riemann invariants of solitonic hydrodynamics}
\label{sec:riemann}
In this section, we provide the derivation of the Riemann invariants
for the soliton train modulation system
\begin{subequations}
  \label{eq:11}
  \begin{align}
    \label{eq:12}
    &\overline{u}_t + V(\overline{u})\overline{u}_x = 0 , \\
    \label{eq:13}
    &a_t +
      c(a,\overline{u}) a_x + f(a,\overline{u}) \overline{u}_x = 0, \\
    \label{eq:5}
    &k_t + \left [ c(a,\ub \right) k ]_x = 0 ,
  \end{align}
\end{subequations}
enabling its reduction to the diagonal form
\begin{subequations}
  \label{eq:14}
  \begin{align}
    \label{eq:15}
    &\ub_t + V(\ub)\ub_x = 0, \\
    \label{eq:16}
    &q_t +  C(q,\ub) q_x = 0 ,\\
    \label{eq:17}
    &(kp)_t +  C(q,\ub) (kp)_x = 0 .
  \end{align}
\end{subequations}
To elucidate the general procedure, we perform explicit calculations
for the KdV equation along with the conduit equation, our primary
example.

First, we notice that equations \eqref{eq:12} and \eqref{eq:13} are
decoupled from \eqref{eq:5}, and have two distinct, real
characteristic velocities $V<c$. This $2 \times 2$ subsystem of
quasi-linear equations is thus strictly hyperbolic and can be
diagonalized for any coupling function $f(a, \ub)$
\cite{whitham_linear_1974}.

The mean field equation is already in diagonal form with the Riemann
invariant $\ub$ associated with the velocity $V$.  The second Riemann
invariant, $q$, depends on $\ub$, $a$ and is associated with the
characteristic velocity $c$.  It can be found by integrating
$f \mathrm{d}\ub + (c-V) \mathrm{d} a$ (see, e.g.,
Ref.~\cite{whitham_linear_1974}). Another way of finding $q$ is to
look for a simple wave relation $a(\ub)$ of the subsystem \eqref{eq:12}
and \eqref{eq:13}.

The coupling function $f(a, \ub)$ is not always readily available, and
its direct computation generally requires the determination of a
singular, soliton limit in the full system of Whitham modulation
equations for the dispersive hydrodynamics
\cite{gurevich_nonlinear_1990}. Below, we use a convenient change of
variables proposed in Ref.~\cite{el_2005} that enables one to
circumvent explicit determination of the coupling function $f$ in the
derivation of the Riemann invariant $q$, utilizing only the known
linear dispersion relation $\omega(k,\overline{u})$ and the soliton
amplitude-speed relation $c(a,\overline{u})$.

Following Ref.~\cite{el_2005}, we introduce a
conjugate (soliton) wavenumber
$\tilde{k} = \tilde{K}(a,\overline{u})$, implicitly determined via the
soliton amplitude-speed relation
\begin{equation}
  \label{eq:6}
  c(a,\overline{u}) =
  \tilde{\omega}(\tilde{k},\overline{u})/\tilde{k},
\end{equation}
where $\tilde{\omega}(\tilde{k},\overline{u}) =
-i\omega(i\tilde{k},\overline{u})$ is the conjugate dispersion, whose
phase velocity coincides with the speed of a soliton.  The conjugate
dispersion relation is realized by linearizing the governing
dispersive hydrodynamic equation, Eq.~(1), with respect to the soliton
solution in the far-field.  Very often, one can explicitly determine
the soliton amplitude-speed relation $c(a,\overline{u})$ hence also
the change of variables $\tilde{k} = \tilde{K}(a,\ub)$ via
Eq.~(\ref{eq:6}).

As a simple example, for the KdV equation we have
$\omega = k \ub - k^3$, $c(a, \ub)=\ub + a/3$, therefore
$\tilde{k}^2 = a/3$.

For the conduit equation, the dispersion and soliton amplitude-speed
relations are
\cite{maiden_observation_2016} 
\begin{equation}
  \label{eq:18}
  \begin{split}
    \omega(k,\overline{u}) &= \frac{2\overline{u} k}{1+\overline{u}k^2}, \\
    c_s(a,\overline{u}) &=
    \frac{\overline{u}}{a^2}
    \{(a+\overline{u})^2[2\ln(1+a/\overline{u})-1]+\overline{u}^2\} .
  \end{split}
\end{equation}
The conjugate wavenumber transformation (\ref{eq:6}) then yields
$\tilde{k}^2 = 1/\overline{u} - 2/c_s(a,\ub)$.  

In the variables $(\tilde{k},\ub)$, simple wave solutions of
Eqs.~\eqref{eq:12} and \eqref{eq:13} satisfy the ordinary differential
equation (ODE) \cite{el_2005}
\begin{equation}
  \label{eq:19}
  \frac{d\tilde{k}}{d \overline{u}} =
  \frac{\tilde{\omega}_{\overline{u}}}{V(\overline{u}) -
    \tilde{\omega}_{\tilde{k}}} .
\end{equation}

\begin{figure*}
  \centering
  \includegraphics[width=18cm]{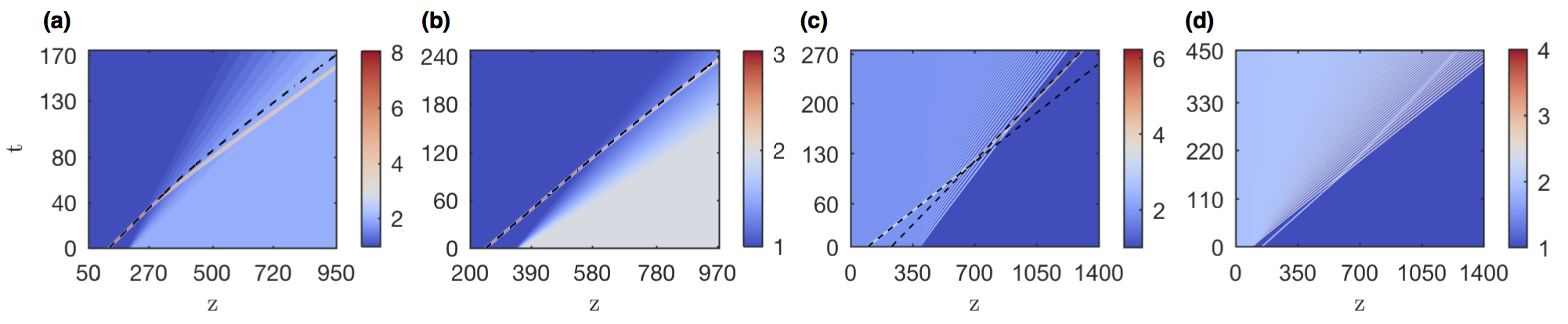}
  \caption{Numerical simulations of the conduit equation (contours)
    and corresponding predicted soliton trajectories (dashed curves).
    a) Soliton-RW transmission for $a_- = 7$, $\ub_- = 1$, and
    $\ub_+ = 2$.  The dashed curve is the integration of
    Eq.~\eqref{eq:14}.  b) Soliton-RW trapping for $a_- = 2$,
    $\ub_- = 1$, and $\ub_+ = 2$.  The dashed curve is the integration
    of Eq.~\eqref{eq:14}.  c) Soliton-DSW transmission for $a_- = 2$,
    $\ub_- = 2$, and $\ub_+ = 1$.  The two dashed curves correspond to
    the predicted soliton trajectory pre and post DSW interaction.
    The post interaction trajectory is determined by the adiabatic
    invariants $q(a,\overline{u})$ and $r(a,\overline{u})$. The
    difference between the $z$ intercepts of these lines is the
    soliton-DSW phase shift.  d) Soliton-DSW trapping for $a_+ = 1.5$,
    $\ub_- = 2$, and $\ub_+ = 1$.}
  \label{fig:kdv_DSW}
\end{figure*}
For the KdV equation, the ODE (\ref{eq:19}) is readily integrated to
yield $2 \ub + 3 \tilde k^2=q$, where $q$ is a constant.

For the conduit equation, integration of (\ref{eq:19}) gives an implicit
determination of $\tilde{k}(\ub)$
\begin{equation}
  \label{eq:20}
  \frac{\ub(2 - \ub \tilde{k}^2)}{(\ub \tilde{k}^2 - 1)^2} = q,
\end{equation}
where $q$, again, is a constant of integration.   

Generally, $q = q(a,\ub)$ is constant along the characteristic
$dx/dt = C(q, \ub)$, where $C(q(a,\ub), \ub) \equiv c(a, \ub)$, and so
is a Riemann invariant satisfying Eq.~\eqref{eq:16}.  For the KdV
equation, we find $q = a + 2\ub$ and $C = (q+\ub)/3$.  For the conduit
equation, we obtain $q(a, \ub) = c(a,\ub)[c(a,\ub)+2\ub]/\ub$ and
$C(q, \ub) = -\ub + \sqrt{\ub(q+\ub)}$ (Eq.~(5)).
 
Having determined two Riemann invariants, $\ub$ and $q(a, \ub)$, of
the system \eqref{eq:11}, the third Riemann invariant, labeled
$r(q, \ub, k)$ and also associated with the characteristic velocity
$C(q, \ub)$, is readily found in the form $r=kp(q,\ub)$ where
$p(q,\ub)$ is given by the general expression (4).

The constancy of $q$ and $r$ lead to the transmission conditions in
Eq.~(8).  These conditions completely determine the
soliton's trajectory post RW or DSW interaction.  They also determine
the conditions for soliton-hydrodynamic trapping.  All four possible
interaction types (soliton-RW, soliton-DSW trapping and transmission)
are shown in the numerical simulations compared with theoretical
predictions in Fig.~\ref{fig:kdv_DSW}.

\section{Appendix B:  Amplitude property of soliton transmission}

\label{sec:trans_p}

Conduit soliton transmission through a RW has the following property.
The amplitude of the transmitted soliton, $a_+$, is always smaller
than the amplitude of the incident soliton, $a_+ < a_-$.  The same
result readily follows for KdV soliton transmission using the first
condition (8) and the expression $q = a + 2 \ub$ for the Riemann
invariant derived in Sec.~\ref{sec:riemann}. We now derive the
extension of this result to the general dispersive hydrodynamic system
Eq.~(1), assuming $V'(u)>0$.

In the mean field, simple wave approximation, the transmission of a
soliton through a RW is determined by the conservation of the second
Riemann invariant, $q(a, \ub)$, of the solitonic hydrodynamic system
\eqref{eq:11} (see the first condition in Eq.~(8)). We are interested
in the sign of the derivative $a_x$ in the course of soliton
transmission. Expressing $a(q, \ub)$, we obtain $a_x=a_{\ub}\ub_x$
(since $q$ is constant). Now, using Eq.~(7), we have
$\ub_x= 1/(t V'(x/t))>0$. Next, $a_{\ub}=-q_{\ub}/q_a$. Hence, $\sgn
\, a_x = -\sgn (q_{\ub} \,q_{a})$ and therefore, $\sgn [a_+ - a_-]=
-\sgn [q_{\ub} \,q_{a}]$ assuming $q_{\ub} \ne 0$, $q_a \ne 0$. Say,
for KdV $q=a + 2 \ub$ so $\sgn [a_+ - a_-]=-1$ as already observed.
By hydrodynamic reciprocity, the amplitude change for soliton-DSW
transmission is opposite to that of soliton-RW transmission.

In a similar manner, the acceleration or deceleration of soliton-RW
transmission can be determined by the positivity or negativity of
$C_x$.  By a similar argument,
$\mathrm{sgn} [C_x] = \mathrm{sgn}[C_{\ub}]$.  For KdV,
$C_{\ub} = 1/3$, and for the conduit equation,
$C_{\ub} = c^2/[2\ub(c+\ub)] > 0$, so a transmitted soliton in both
cases is accelerated by the RW.  By hydrodynamic reciprocity, a
soliton is decelerated by a DSW.

\section{Appendix C:  Fluid Conduit Experimental Details}
\label{sec:fluid-cond-exper}

The experimental setup is similar to that described in
\cite{maiden_observation_2016} and consists of a square
acrylic column with dimensions $5$ $\times$ $5$ $\times$ $183$ cm$^3$.
The column is filled with glycerol, a highly viscous, transparent,
exterior fluid.  A nozzle installed at the base of the column allows
for the injection of the interior fluid. To eliminate surface tension
effects, the interior fluid is a miscible solution of glycerol, water,
and black food coloring.  As a result, the interior fluid has both
lower density and viscosity than the exterior fluid.  The properties
of the interior and exterior fluids are dynamic viscosity $\mu^{(i)}=
51\pm 1$ cP, $\mu^{(e)}=1200 \pm 20$ cP and density $\rho^{(i)}=1.2286
\pm 0.0001$ g/cm$^3$, and $\rho^{(e)}=1.2587 \pm 0.0001$ g/cm$^3$ with
a superscript denoting \textit{i}nterior/\textit{e}xterior.

Interior fluid is drawn from a reservoir and injected through the
nozzle via a computer controlled high precision piston pump.  By
injecting at a constant rate, the buoyant interior fluid establishes a
stable, vertically uniform fluid conduit. By varying the injection
rate in a precise manner, conduit solitons, RWs, and DSWs can be
reliably generated.  For each trial, a volumetric flow rate profile
$Q(t)$ is generated based on these fluid properties that results in a
long box followed by a soliton of chosen amplitude. The smaller
conduit diameter is set by the background flow rate $Q = 0.25$ mL/min.
The maximum flow rate for the long box is $Q = 0.7656$ mL/min. These
two flow rates correspond to a nondimensional jump in cross-sectional
area from 1 to 1.75 that defines $\overline{u}_\pm$.  The box is
sufficiently long that the trailing edge acts as a RW and leading edge
acts as a DSW at the time of their respective interactions with the
soliton.

The cross-sectional area of the fluid conduit corresponds to the
dispersive hydrodynamic medium of interest.  Data acquisition of
$\ub_\pm$ and $a_\pm$ is performed using three high resolution
cameras, two equipped with macro lenses and one with a zoom lens. The
macro lens cameras are near the bottom and top of the conduit, and the
zoom lens near the middle, for extracting precise conduit diameter and
soliton amplitude information. The cameras take several
high-resolution images of the soliton as it passes through their
respective viewscreens, as well as pictures of the background conduit
before and after the hydrodynamic structure has passed. Correction for
the refractive index of glycerin is calibrated via images of a
cylinder of known height and width dropped into the center of the
apparatus before the experiment.
    
The camera images are processed in \textsc{Matlab} to extract the
conduit edges by taking a horizontal row of pixels and calculating the
maximum and minimum derivatives for each row.  As the background is
white and the conduit is black, this finds the approximate boundary.
The conduit's slight variability from vertical gets enhanced during
image processing so we use a background subtraction method to extract
the measured non-dimensional conduit area from the images reported in
Figs.~2 and 4.  Prior to inducing interfacial dynamics, we take ten
images, extract non-dimensional edge data, average all the edges and
subtract this from edges extracted for the trial.  The data is then
sent through a low-pass filter to reduce noise from the pixelation of
the photograph and any impurities (such as bubbles) in the exterior
fluid.  After converting to area and rescaling the smaller background
area to unity, we then determine the soliton amplitudes before and
after tunneling.  Calculations suggest that density variations of 1\%
in the exterior fluid can lead to a 10\% change in the background
conduit diameter.  We observe an increase in the conduit diameter for
the top camera, relative to the bottom and middle cameras by 10.1\%,
which we attribute to density variation of the external fluid.
Because the model assumes no density variation, we accommodate this
discrepancy by scaling all amplitude measurements from the top camera
by the factor $1.101^2 = 1.212$.

\begin{acknowledgments}
  This work was supported by NSF CAREER DMS-1255422 (DVA, NAF, MAH),
  the NSF GRFP (MDM), NSF EXTREEMS-QED DMS-1407340 (DVA), and EPSRC
  grant EP/R00515X/1 (GAE).  GAE and MAH gratefully acknowledge the
  London Mathematical Society for supporting a Research in Pairs
  visit.
\end{acknowledgments}


\begin{thebibliography}{33}%
\makeatletter
\providecommand \@ifxundefined [1]{%
 \@ifx{#1\undefined}
}%
\providecommand \@ifnum [1]{%
 \ifnum #1\expandafter \@firstoftwo
 \else \expandafter \@secondoftwo
 \fi
}%
\providecommand \@ifx [1]{%
 \ifx #1\expandafter \@firstoftwo
 \else \expandafter \@secondoftwo
 \fi
}%
\providecommand \natexlab [1]{#1}%
\providecommand \enquote  [1]{``#1''}%
\providecommand \bibnamefont  [1]{#1}%
\providecommand \bibfnamefont [1]{#1}%
\providecommand \citenamefont [1]{#1}%
\providecommand \href@noop [0]{\@secondoftwo}%
\providecommand \href [0]{\begingroup \@sanitize@url \@href}%
\providecommand \@href[1]{\@@startlink{#1}\@@href}%
\providecommand \@@href[1]{\endgroup#1\@@endlink}%
\providecommand \@sanitize@url [0]{\catcode `\\12\catcode `\$12\catcode
  `\&12\catcode `\#12\catcode `\^12\catcode `\_12\catcode `\%12\relax}%
\providecommand \@@startlink[1]{}%
\providecommand \@@endlink[0]{}%
\providecommand \url  [0]{\begingroup\@sanitize@url \@url }%
\providecommand \@url [1]{\endgroup\@href {#1}{\urlprefix }}%
\providecommand \urlprefix  [0]{URL }%
\providecommand \Eprint [0]{\href }%
\providecommand \doibase [0]{http://dx.doi.org/}%
\providecommand \selectlanguage [0]{\@gobble}%
\providecommand \bibinfo  [0]{\@secondoftwo}%
\providecommand \bibfield  [0]{\@secondoftwo}%
\providecommand \translation [1]{[#1]}%
\providecommand \BibitemOpen [0]{}%
\providecommand \bibitemStop [0]{}%
\providecommand \bibitemNoStop [0]{.\EOS\space}%
\providecommand \EOS [0]{\spacefactor3000\relax}%
\providecommand \BibitemShut  [1]{\csname bibitem#1\endcsname}%
\let\auto@bib@innerbib\@empty
\bibitem [{\citenamefont {Landau}\ and\ \citenamefont
  {Lifshitz}(1987)}]{landau_fluid_1987}%
  \BibitemOpen
  \bibfield  {author} {\bibinfo {author} {\bibfnamefont {L.~D.}\ \bibnamefont
  {Landau}}\ and\ \bibinfo {author} {\bibfnamefont {E.}~\bibnamefont
  {Lifshitz}},\ }\href@noop {} {\emph {\bibinfo {title} {Fluid {Mechanics}}}},\
  \bibinfo {edition} {2nd}\ ed.\ (\bibinfo  {publisher}
  {Butterworth-Heinemann},\ \bibinfo {year} {1987})\BibitemShut {NoStop}%
\bibitem [{\citenamefont {Carusotto}\ and\ \citenamefont
  {Ciuti}(2013)}]{carusotto_quantum_2013}%
  \BibitemOpen
  \bibfield  {author} {\bibinfo {author} {\bibfnamefont {I.}~\bibnamefont
  {Carusotto}}\ and\ \bibinfo {author} {\bibfnamefont {C.}~\bibnamefont
  {Ciuti}},\ }\href@noop {} {\bibfield  {journal} {\bibinfo  {journal} {Rev.
  Mod. Phys.}\ }\textbf {\bibinfo {volume} {85}},\ \bibinfo {pages} {299}
  (\bibinfo {year} {2013})}\BibitemShut {NoStop}%
\bibitem [{\citenamefont {Fradkin}(2013)}]{fradkin_field_2013}%
  \BibitemOpen
  \bibfield  {author} {\bibinfo {author} {\bibfnamefont {E.}~\bibnamefont
  {Fradkin}},\ }\href@noop {} {\emph {\bibinfo {title} {Field {Theories} of
  {Condensed} {Matter} {Physics}}}},\ \bibinfo {edition} {2nd}\ ed.\ (\bibinfo
  {publisher} {Cambridge University Press},\ \bibinfo {address} {Cambridge,
  UK},\ \bibinfo {year} {2013})\BibitemShut {NoStop}%
\bibitem [{\citenamefont {Wyatt}\ and\ \citenamefont
  {Trahan}(2005)}]{wyatt_quantum_2005}%
  \BibitemOpen
  \bibfield  {author} {\bibinfo {author} {\bibfnamefont {R.~E.}\ \bibnamefont
  {Wyatt}}\ and\ \bibinfo {author} {\bibfnamefont {C.~J.}\ \bibnamefont
  {Trahan}},\ }\href@noop {} {\emph {\bibinfo {title} {Quantum dynamics with
  trajectories: introduction to quantum hydrodynamics}}},\ \bibinfo {series}
  {Interdisciplinary applied mathematics}\ \bibinfo {number} {v. 28}\
  (\bibinfo  {publisher} {Springer},\ \bibinfo {address} {New York},\ \bibinfo
  {year} {2005})\BibitemShut {NoStop}%
\bibitem [{\citenamefont {Whitham}(1974)}]{whitham_linear_1974}%
  \BibitemOpen
  \bibfield  {author} {\bibinfo {author} {\bibfnamefont {G.~B.}\ \bibnamefont
  {Whitham}},\ }\href@noop {} {\emph {\bibinfo {title} {Linear and nonlinear
  waves}}}\ (\bibinfo  {publisher} {Wiley},\ \bibinfo {address} {New York},\
  \bibinfo {year} {1974})\BibitemShut {NoStop}%
\bibitem [{\citenamefont {El}\ and\ \citenamefont
  {Hoefer}(2016)}]{el_dispersive_2016}%
  \BibitemOpen
  \bibfield  {author} {\bibinfo {author} {\bibfnamefont {G.~A.}\ \bibnamefont
  {El}}\ and\ \bibinfo {author} {\bibfnamefont {M.~A.}\ \bibnamefont
  {Hoefer}},\ }\href@noop {} {\bibfield  {journal} {\bibinfo  {journal}
  {Physica D}\ }\textbf {\bibinfo {volume} {333}},\ \bibinfo {pages} {11}
  (\bibinfo {year} {2016})}\BibitemShut {NoStop}%
\bibitem [{\citenamefont {Gurevich}\ and\ \citenamefont
  {Pitaevskii}(1974)}]{gurevich_nonstationary_1974}%
  \BibitemOpen
  \bibfield  {author} {\bibinfo {author} {\bibfnamefont {A.~V.}\ \bibnamefont
  {Gurevich}}\ and\ \bibinfo {author} {\bibfnamefont {L.~P.}\ \bibnamefont
  {Pitaevskii}},\ }\href@noop {} {\bibfield  {journal} {\bibinfo  {journal}
  {Sov. Phys. JETP}\ }\textbf {\bibinfo {volume} {38}},\ \bibinfo {pages} {291}
  (\bibinfo {year} {1974})},\ \bibinfo {note} {translation from Russian of A. V.
  Gurevich and L. P. Pitaevskii, Zh. Eksp. Teor. Fiz. 65, 590-604 (August
  1973).}\BibitemShut {Stop}%
\bibitem [{\citenamefont {Hoefer}\ \emph {et~al.}(2006)\citenamefont {Hoefer},
  \citenamefont {Ablowitz}, \citenamefont {Coddington}, \citenamefont
  {Cornell}, \citenamefont {Engels},\ and\ \citenamefont
  {Schweikhard}}]{hoefer_dispersive_2006-1}%
  \BibitemOpen
  \bibfield  {author} {\bibinfo {author} {\bibfnamefont {M.~A.}\ \bibnamefont
  {Hoefer}}, \bibinfo {author} {\bibfnamefont {M.~J.}\ \bibnamefont
  {Ablowitz}}, \bibinfo {author} {\bibfnamefont {I.}~\bibnamefont
  {Coddington}}, \bibinfo {author} {\bibfnamefont {E.~A.}\ \bibnamefont
  {Cornell}}, \bibinfo {author} {\bibfnamefont {P.}~\bibnamefont {Engels}}, \
  and\ \bibinfo {author} {\bibfnamefont {V.}~\bibnamefont {Schweikhard}},\
  }\href@noop {} {\bibfield  {journal} {\bibinfo  {journal} {Phys. Rev. A}\
  }\textbf {\bibinfo {volume} {74}},\ \bibinfo {pages} {023623} (\bibinfo
  {year} {2006})}\BibitemShut {NoStop}%
\bibitem [{\citenamefont {Mo}\ \emph {et~al.}(2013)\citenamefont {Mo},
  \citenamefont {Kishek}, \citenamefont {Feldman}, \citenamefont {Haber},
  \citenamefont {Beaudoin}, \citenamefont {O’Shea},\ and\ \citenamefont
  {Thangaraj}}]{mo_experimental_2013}%
  \BibitemOpen
  \bibfield  {author} {\bibinfo {author} {\bibfnamefont {Y.~C.}\ \bibnamefont
  {Mo}}, \bibinfo {author} {\bibfnamefont {R.~A.}\ \bibnamefont {Kishek}},
  \bibinfo {author} {\bibfnamefont {D.}~\bibnamefont {Feldman}}, \bibinfo
  {author} {\bibfnamefont {I.}~\bibnamefont {Haber}}, \bibinfo {author}
  {\bibfnamefont {B.}~\bibnamefont {Beaudoin}}, \bibinfo {author}
  {\bibfnamefont {P.~G.}\ \bibnamefont {O’Shea}}, \ and\ \bibinfo {author}
  {\bibfnamefont {J.~C.~T.}\ \bibnamefont {Thangaraj}},\ }\href@noop {}
  {\bibfield  {journal} {\bibinfo  {journal} {Phys. Rev. Lett.}\ }\textbf
  {\bibinfo {volume} {110}},\ \bibinfo {pages} {084802} (\bibinfo {year}
  {2013})}\BibitemShut {NoStop}%
\bibitem [{\citenamefont {Wan}\ \emph {et~al.}(2007)\citenamefont {Wan},
  \citenamefont {Jia},\ and\ \citenamefont {Fleischer}}]{wan_dispersive_2007}%
  \BibitemOpen
  \bibfield  {author} {\bibinfo {author} {\bibfnamefont {W.}~\bibnamefont
  {Wan}}, \bibinfo {author} {\bibfnamefont {S.}~\bibnamefont {Jia}}, \ and\
  \bibinfo {author} {\bibfnamefont {J.~W.}\ \bibnamefont {Fleischer}},\
  }\href@noop {} {\bibfield  {journal} {\bibinfo  {journal} {Nat. Phys.}\
  }\textbf {\bibinfo {volume} {3}},\ \bibinfo {pages} {46} (\bibinfo {year}
  {2007})}\BibitemShut {NoStop}%
\bibitem [{\citenamefont {Xu}\ \emph {et~al.}(2017)\citenamefont {Xu},
  \citenamefont {Conforti}, \citenamefont {Kudlinski}, \citenamefont {Mussot},\
  and\ \citenamefont {Trillo}}]{xu_dispersive_2017-1}%
  \BibitemOpen
  \bibfield  {author} {\bibinfo {author} {\bibfnamefont {G.}~\bibnamefont
  {Xu}}, \bibinfo {author} {\bibfnamefont {M.}~\bibnamefont {Conforti}},
  \bibinfo {author} {\bibfnamefont {A.}~\bibnamefont {Kudlinski}}, \bibinfo
  {author} {\bibfnamefont {A.}~\bibnamefont {Mussot}}, \ and\ \bibinfo {author}
  {\bibfnamefont {S.}~\bibnamefont {Trillo}},\ }\href@noop {} {\bibfield
  {journal} {\bibinfo  {journal} {Phys. Rev. Lett.}\ }\textbf {\bibinfo {volume}
  {118}},\ \bibinfo {pages} {254101} (\bibinfo {year} {2017})}\BibitemShut
  {NoStop}%
\bibitem [{\citenamefont {Trillo}\ \emph {et~al.}(2016)\citenamefont {Trillo},
  \citenamefont {Klein}, \citenamefont {Clauss},\ and\ \citenamefont
  {Onorato}}]{trillo_observation_2016}%
  \BibitemOpen
  \bibfield  {author} {\bibinfo {author} {\bibfnamefont {S.}~\bibnamefont
  {Trillo}}, \bibinfo {author} {\bibfnamefont {M.}~\bibnamefont {Klein}},
  \bibinfo {author} {\bibfnamefont {G.}~\bibnamefont {Clauss}}, \ and\ \bibinfo
  {author} {\bibfnamefont {M.}~\bibnamefont {Onorato}},\ }\href@noop {}
  {\bibfield  {journal} {\bibinfo  {journal} {Physica D}\ }\textbf {\bibinfo
  {volume} {333}},\ \bibinfo {pages} {276} (\bibinfo {year}
  {2016})}\BibitemShut {NoStop}%
\bibitem [{\citenamefont {Maiden}\ \emph {et~al.}(2016)\citenamefont {Maiden},
  \citenamefont {Lowman}, \citenamefont {Anderson}, \citenamefont {Schubert},\
  and\ \citenamefont {Hoefer}}]{maiden_observation_2016}%
  \BibitemOpen
  \bibfield  {author} {\bibinfo {author} {\bibfnamefont {M.~D.}\ \bibnamefont
  {Maiden}}, \bibinfo {author} {\bibfnamefont {N.~K.}\ \bibnamefont {Lowman}},
  \bibinfo {author} {\bibfnamefont {D.~V.}\ \bibnamefont {Anderson}}, \bibinfo
  {author} {\bibfnamefont {M.~E.}\ \bibnamefont {Schubert}}, \ and\ \bibinfo
  {author} {\bibfnamefont {M.~A.}\ \bibnamefont {Hoefer}},\ }\href@noop {}
  {\bibfield  {journal} {\bibinfo  {journal} {Phys. Rev. Lett.}\ }\textbf
  {\bibinfo {volume} {116}},\ \bibinfo {pages} {174501} (\bibinfo {year}
  {2016})}\BibitemShut {NoStop}%
\bibitem [{\citenamefont {Janantha}\ \emph {et~al.}(2017)\citenamefont
    {Janantha}, \citenamefont {Sprenger}, \citenamefont {Hoefer},\
    and\ \citenamefont {Wu}}]{janantha_observation_2017}%
  \BibitemOpen \bibfield {author} {\bibinfo {author} {\bibfnamefont
      {P.~A.~Praveen}\ \bibnamefont {Janantha}}, \bibinfo {author}
    {\bibfnamefont {P.}~\bibnamefont {Sprenger}}, \bibinfo {author}
    {\bibfnamefont {M.}~\bibnamefont {Hoefer}}, \ and\ \bibinfo
    {author} {\bibfnamefont {M.}~\bibnamefont {Wu}},\ }\href@noop {}
  {\bibfield {journal} {\bibinfo {journal} {Phys. Rev. Lett.}\
    }\textbf {\bibinfo {volume} {119}},\ \bibinfo {pages} {024101}
    (\bibinfo {year} {2017})}\BibitemShut {NoStop}%
\bibitem [{\citenamefont {Zabusky}\ and\ \citenamefont
  {Kruskal}(1965)}]{zabusky_interaction_1965}%
  \BibitemOpen
  \bibfield  {author} {\bibinfo {author} {\bibfnamefont {N.~J.}\ \bibnamefont
  {Zabusky}}\ and\ \bibinfo {author} {\bibfnamefont {M.~D.}\ \bibnamefont
  {Kruskal}},\ }\href@noop {} {\bibfield  {journal} {\bibinfo  {journal} {Phys.
  Rev. Lett.}\ }\textbf {\bibinfo {volume} {15}},\ \bibinfo {pages} {240}
  (\bibinfo {year} {1965})}\BibitemShut {NoStop}%
\bibitem [{\citenamefont {Drazin}\ and\ \citenamefont
  {Johnson}(1989)}]{drazin_solitons:_1989}%
  \BibitemOpen
  \bibfield  {author} {\bibinfo {author} {\bibfnamefont {P.~G.}\ \bibnamefont
  {Drazin}}\ and\ \bibinfo {author} {\bibfnamefont {R.~S.}\ \bibnamefont
  {Johnson}},\ }\href@noop {} {\emph {\bibinfo {title} {Solitons: an
  introduction}}}\ (\bibinfo  {publisher} {Cambridge University Press},\
  \bibinfo {address} {Cambridge, UK},\ \bibinfo {year} {1989})\BibitemShut
  {NoStop}%
\bibitem [{\citenamefont {Remoissenet}(2013)}]{remoissenet_waves_2013}%
  \BibitemOpen
  \bibfield  {author} {\bibinfo {author} {\bibfnamefont {M.}~\bibnamefont
  {Remoissenet}},\ }\href@noop {} {\emph {\bibinfo {title} {Waves called
  solitons: concepts and experiments}}},\ \bibinfo {edition} {3rd}\ ed.\
  (\bibinfo  {publisher} {Springer},\ \bibinfo {address} {New York},\ \bibinfo
  {year} {2013})\BibitemShut {NoStop}%
\bibitem [{\citenamefont {Lowman}\ and\ \citenamefont
  {Hoefer}(2013{\natexlab{a}})}]{lowman_dispersive_2013-1}%
  \BibitemOpen
  \bibfield  {author} {\bibinfo {author} {\bibfnamefont {N.~K.}\ \bibnamefont
  {Lowman}}\ and\ \bibinfo {author} {\bibfnamefont {M.~A.}\ \bibnamefont
  {Hoefer}},\ }\href@noop {} {\bibfield  {journal} {\bibinfo  {journal} {Phys.
  Rev. E}\ }\textbf {\bibinfo {volume} {88}},\ \bibinfo {pages} {023016}
  (\bibinfo {year} {2013}{\natexlab{a}})}\BibitemShut {NoStop}%
\bibitem [{\citenamefont {Olson}\ and\ \citenamefont
  {Christensen}(1986)}]{olson_solitary_1986}%
  \BibitemOpen
  \bibfield  {author} {\bibinfo {author} {\bibfnamefont {P.}~\bibnamefont
  {Olson}}\ and\ \bibinfo {author} {\bibfnamefont {U.}~\bibnamefont
  {Christensen}},\ }\href@noop {} {\bibfield  {journal} {\bibinfo  {journal} {J.
  Geophys. Res.}\ }\textbf {\bibinfo {volume} {91}},\ \bibinfo {pages} {6367}
  (\bibinfo {year} {1986})}\BibitemShut {NoStop}%
\bibitem [{\citenamefont {Scott}\ \emph {et~al.}(1986)\citenamefont {Scott},
  \citenamefont {Stevenson},\ and\ \citenamefont
  {Whitehead}}]{scott_observations_1986}%
  \BibitemOpen
  \bibfield  {author} {\bibinfo {author} {\bibfnamefont {D.~R.}\ \bibnamefont
  {Scott}}, \bibinfo {author} {\bibfnamefont {D.~J.}\ \bibnamefont
  {Stevenson}}, \ and\ \bibinfo {author} {\bibfnamefont {J.~A.}\ \bibnamefont
  {Whitehead}},\ }\href@noop {} {\bibfield  {journal} {\bibinfo  {journal}
  {Nature}\ }\textbf {\bibinfo {volume} {319}},\ \bibinfo {pages} {759}
  (\bibinfo {year} {1986})}\BibitemShut {NoStop}%
\bibitem [{\citenamefont {Lowman}\ \emph {et~al.}(2014)\citenamefont {Lowman},
  \citenamefont {Hoefer},\ and\ \citenamefont {El}}]{lowman_interactions_2014}%
  \BibitemOpen
  \bibfield  {author} {\bibinfo {author} {\bibfnamefont {N.~K.}\ \bibnamefont
  {Lowman}}, \bibinfo {author} {\bibfnamefont {M.~A.}\ \bibnamefont {Hoefer}},
  \ and\ \bibinfo {author} {\bibfnamefont {G.~A.}\ \bibnamefont {El}},\
  }\href@noop {} {\bibfield  {journal} {\bibinfo  {journal} {J. Fluid Mech.}\
  }\textbf {\bibinfo {volume} {750}},\ \bibinfo {pages} {372} (\bibinfo {year}
  {2014})}\BibitemShut {NoStop}%
\bibitem [{\citenamefont {Selvam}\ \emph {et~al.}(2009)\citenamefont {Selvam},
  \citenamefont {Talon}, \citenamefont {Lesshafft},\ and\ \citenamefont
  {Meiburg}}]{selvam_convective_2009}%
  \BibitemOpen
  \bibfield  {author} {\bibinfo {author} {\bibfnamefont {B.}~\bibnamefont
  {Selvam}}, \bibinfo {author} {\bibfnamefont {L.}~\bibnamefont {Talon}},
  \bibinfo {author} {\bibfnamefont {L.}~\bibnamefont {Lesshafft}}, \ and\
  \bibinfo {author} {\bibfnamefont {E.}~\bibnamefont {Meiburg}},\ }\href@noop
  {} {\bibfield  {journal} {\bibinfo  {journal} {J. Fluid Mech.}\ }\textbf
  {\bibinfo {volume} {618}},\ \bibinfo {pages} {323} (\bibinfo {year}
  {2009})}\BibitemShut {NoStop}%
\bibitem [{\citenamefont {El}\ \emph {et~al.}(2017)\citenamefont {El},
  \citenamefont {Hoefer},\ and\ \citenamefont {Shearer}}]{el_dispersive_2017}%
  \BibitemOpen
  \bibfield  {author} {\bibinfo {author} {\bibfnamefont {G.}~\bibnamefont
  {El}}, \bibinfo {author} {\bibfnamefont {M.}~\bibnamefont {Hoefer}}, \ and\
  \bibinfo {author} {\bibfnamefont {M.}~\bibnamefont {Shearer}},\ }\href@noop
  {} {\bibfield  {journal} {\bibinfo  {journal} {SIAM Rev.}\ }\textbf {\bibinfo
  {volume} {59}},\ \bibinfo {pages} {3} (\bibinfo {year} {2017})}\BibitemShut
  {NoStop}%
\bibitem [{\citenamefont {Grimshaw}(1979)}]{grimshaw_slowly_1979}%
  \BibitemOpen
  \bibfield  {author} {\bibinfo {author} {\bibfnamefont {R.}~\bibnamefont
  {Grimshaw}},\ }\href@noop {} {\bibfield  {journal} {\bibinfo  {journal}
  {Proc. R. Soc. Lond. A}\ }\textbf {\bibinfo {volume} {368}},\ \bibinfo
  {pages} {359} (\bibinfo {year} {1979})}\BibitemShut {NoStop}%
\bibitem [{\citenamefont {Gurevich}\ \emph {et~al.}(1990)\citenamefont
    {Gurevich}, \citenamefont {Krylov},\ and\ \citenamefont
    {El}}]{gurevich_nonlinear_1990}%
  \BibitemOpen \bibfield {author} {\bibinfo {author} {\bibfnamefont
      {A.~V.}\ \bibnamefont {Gurevich}}, \bibinfo {author}
    {\bibfnamefont {A.~L.}\ \bibnamefont {Krylov}}, \ and\ \bibinfo
    {author} {\bibfnamefont {G.~A.}\ \bibnamefont {El}},\ }\href@noop
  {} {\bibfield {journal} {\bibinfo {journal} {Sov. Phys.  JETP}\
    }\textbf {\bibinfo {volume} {71}},\ \bibinfo {pages} {899}
    (\bibinfo {year} {1990})}\BibitemShut {NoStop}%
\bibitem [{\citenamefont {El}\ \emph {et~al.}(2007)\citenamefont
    {El}, \citenamefont {Grimshaw},\ and\ \citenamefont
  {Kamchatnov}}]{el_evolution_2007}%
\BibitemOpen
\bibfield  {author} {\bibinfo {author} {\bibfnamefont {G.~A.}\ \bibnamefont
  {El}}, \bibinfo {author} {\bibfnamefont {R.~H.~J.}\ \bibnamefont
  {Grimshaw}}, \ and\ \bibinfo {author} {\bibfnamefont {A.~M.}\ \bibnamefont
  {Kamchatnov}},\ }\href@noop {} {\bibfield  {journal} {\bibinfo
  {journal} {J. Fluid Mech.}\ }\textbf {\bibinfo {volume} {585}},\ \bibinfo {pages} {213} (\bibinfo
  {year} {2007})}\BibitemShut {NoStop}%
\bibitem [{\citenamefont {El}\ \emph {et~al.}(2012)\citenamefont
    {El}, \citenamefont {Grimshaw},\ and\ \citenamefont
  {Tiong}}]{el_transformation_2012}%
\BibitemOpen
\bibfield  {author} {\bibinfo {author} {\bibfnamefont {G.~A.}\ \bibnamefont
  {El}}, \bibinfo {author} {\bibfnamefont {R.~H.~J.}\ \bibnamefont
  {Grimshaw}}, \ and\ \bibinfo {author} {\bibfnamefont {W.~K.}\ \bibnamefont
  {Tiong}},\ }\href@noop {} {\bibfield  {journal} {\bibinfo
  {journal} {J. Fluid Mech.}\ }\textbf {\bibinfo {volume} {709}},\ \bibinfo {pages} {371} (\bibinfo
  {year} {2012})}\BibitemShut {NoStop}%
\bibitem [{\citenamefont {Grimshaw}\ \emph {et~al.}(2016)\citenamefont
    {Grimshaw}\ and\ \citenamefont
  {Yuan}}]{grimshaw_propagation_2016}%
\BibitemOpen
\bibfield  {author} {\bibinfo {author} {\bibfnamefont {R.}\ \bibnamefont
  {Grimshaw}}\ and\ \bibinfo {author} {\bibfnamefont {C.}\ \bibnamefont
  {Yuan}},\ }\href@noop {} {\bibfield  {journal} {\bibinfo
  {journal} {Physica D}\ }\textbf {\bibinfo {volume} {333}},\ \bibinfo {pages} {200} (\bibinfo
  {year} {2016})}\BibitemShut {NoStop}%
\bibitem [{\citenamefont {Grimshaw}\ \emph {et~al.}(2016)\citenamefont
    {Yuan}, \citenamefont {Grimshaw} and\ \citenamefont
  {Yuan}}]{grimshaw_depression_2016}%
\BibitemOpen
\bibfield  {author} {\bibinfo {author} {\bibfnamefont {R.}\ \bibnamefont
  {Grimshaw}}\ and\ \bibinfo {author} {\bibfnamefont {C.}\ \bibnamefont
  {Yuan}},\ }\href@noop {} {\bibfield  {journal} {\bibinfo
  {journal} {Nat. Hazards}\ }\textbf {\bibinfo {volume} {84}},\ \bibinfo {pages} {S493} (\bibinfo
  {year} {2016})}\BibitemShut {NoStop}%
\bibitem [{\citenamefont {Kivshar}\ and\ \citenamefont
  {Malomed}(1989)}]{kivshar_dynamics_1989}%
  \BibitemOpen
  \bibfield  {author} {\bibinfo {author} {\bibfnamefont {Y.~S.}\ \bibnamefont
  {Kivshar}}\ and\ \bibinfo {author} {\bibfnamefont {B.~A.}\ \bibnamefont
  {Malomed}},\ }\href@noop {} {\bibfield  {journal} {\bibinfo  {journal} {Rev.
  Mod. Phys.}\ }\textbf {\bibinfo {volume} {61}},\ \bibinfo {pages} {763}
  (\bibinfo {year} {1989})}\BibitemShut {NoStop}%
\bibitem [{\citenamefont {Lowman}\ and\ \citenamefont
  {Hoefer}(2013{\natexlab{b}})}]{lowman_dispersive_2013}%
  \BibitemOpen
  \bibfield  {author} {\bibinfo {author} {\bibfnamefont {N.~K.}\ \bibnamefont
  {Lowman}}\ and\ \bibinfo {author} {\bibfnamefont {M.~A.}\ \bibnamefont
  {Hoefer}},\ }\href@noop {} {\bibfield  {journal} {\bibinfo  {journal} {J.
  Fluid Mech.}\ }\textbf {\bibinfo {volume} {718}},\ \bibinfo {pages} {524}
  (\bibinfo {year} {2013}{\natexlab{b}})}\BibitemShut {NoStop}%
\bibitem [{\citenamefont {Sprenger}(2018)}]{sprenger_soliton_2018}%
  \BibitemOpen
  \bibfield  {author} {\bibinfo {author} {\bibfnamefont {P.}\ \bibnamefont
  {Sprenger}}, \bibinfo {author} {\bibfnamefont {M.~A.}\ \bibnamefont
  {Hoefer}},\ and\ \bibinfo {author} {\bibnamefont {G.~A.}\
  \bibnamefont {El}},\ }\href@noop {} {\bibfield  {journal} {\bibinfo
  {journal} {preprint arXiv:1711.05239}\ }
  (\bibinfo {year} {2018})}\BibitemShut {NoStop}%
\bibitem [{\citenamefont {El}(2005)}]{el_2005}%
  \BibitemOpen
  \bibfield  {author} {\bibinfo {author} {\bibfnamefont {G.~A.}~\bibnamefont
  {El}},\ }\href@noop {} {\bibfield  {journal} {\bibinfo  {journal}
  {Chaos}\ }\textbf {\bibinfo {volume} {15}},\ \bibinfo
  {pages} {037103} (\bibinfo {year} {2005})}\BibitemShut {NoStop}%
\end{thebibliography}

%

\end{document}